\documentclass[12pt,a4paper]{article}
\usepackage[cp1251]{inputenc}
\usepackage{floatflt}
\usepackage[T1]{fontenc}
\usepackage{amssymb,amsfonts,amsmath}
\usepackage[russian,english]{babel}
\usepackage{hyperref}
\usepackage[dvips]{graphicx}

\newtheorem{theorem}{Theorem}
\newtheorem{corollary}{Corollary}

\author{ A. A. Magazev{\footnote {Omsk State University, Omsk, Russia, e-mail: magaz@phys.omsu.omskreg.ru} }, I. V. Shirokov {\footnote {Omsk State University, Omsk, Russia, e-mail: shirokov@univer.omsk.su}}}
\title{Integration of geodesic flows on homogeneous
spaces: the case of a wild lie group}
\date{}

\begin{document}
\maketitle

\begin{abstract}
We obtain necessary and sufficient conditions for the integrability in quadratures of
geodesic flows on homogeneous spaces $M$ with invariant and central metrics. The proposed
integration algorithm consists in using a special canonical transformation in the space
$T^*M$ based on constructing the canonical coordinates on the orbits of the coadjoint
representation and on the simplectic sheets of the Poisson algebra of invariant
functions. This algorithm is applicable to integrating geodesic flows on homogeneous
spaces of a wild Lie group.
\end{abstract}

\section{Introduction}

This paper is devoted to the problem of integrating geodesic flows on homogeneous spaces.
We consider two classes of metrics on a homogeneous right $G$-space $M = G/H$ connected
with the transformation group, namely, the $G$-invariant metrics and the so-called
central metrics, whose definition is given in Sec. 2. In this paper, we obtain necessary
and sufficient conditions for the integrability in quadratures of geodesic flows for such
metrics. These conditions are expressed in terms of nonnegative integers (the degree of
degeneracy, index, and defect of the homogeneous space) that can be easily found from the
well-known structure constants of the algebra $G$ of the group $G$ and subalgebra $H$ of
the isotropy group $H$. We note that the spaces with zero defect are commutative spaces
whose special cases are symmetric and weakly symmetric spaces.

The algorithm for integrating geodesic flows consists in using a special canonical
transformation in the space $T^*M$ based on constructing the canonical Darboux
coordinates on the orbits of the coadjoint representation and on the simplectic sheets of
the Poisson algebra of invariant functions. In the trivial case of a commutative group
$G$, this canonical transformation consists in the transition to the "action–angle"
variables. As we show in this paper, the proposed integration algorithm is especially
useful in the case where $G$ is a wild Lie group with a non-Hausdorff orbit space.

The study of wild Lie groups is of considerable interest for group representation theory
and mathematical physics. The characteristic feature of these groups is that there are
several nonequivalent ways of decomposing their regular representations into irreducible
components. The Auslander–Kostant theorem \cite{1} gives a criterion for assessing
whether a simply connected solvable Lie group belongs to the class of wild Lie groups in
terms of the orbits of the coadjoint representation (K-orbits). In particular, a Lie
group is wild if the Kirillov simplectic form on a K-orbit is not exact or if the space
of the K-orbits of this group is nonsemiseparated (and, as a consequence, non-Hausdorff).
Examples of such groups are given in \cite{2}.

In this work, we are interested in wild Lie groups with a non-Hausdorff orbit space. This
interest is mainly because the non-Hausdorff character of the space of K-orbits becomes
important when geodesic flows are integrated on homogeneous spaces endowed by the metrics
connected with this group, namely, by $G$-invariant and central metrics. In these cases,
geodesic flows allow for an infinite-valued integral of motion whose values form a dense
subset of the real axis or an open segment. It is clear that level surfaces of such an
integral are not well defined and the standard methods of integrating Hamiltonian systems
are inapplicable in this case.

We consider the five-dimensional solvable wild Lie group with a nonsemiseparated space of
K-orbits described in \cite{2}. As shown in the present work, there are three
nonequivalent four-dimensional homogeneous G-spaces and the group G acts effectively on
only one of them. Passing to the canonical coordinates on a K-orbit \cite{3}, we
integrate the geodesic flow on this homogeneous space endowed by the central metric. This
paper is essentially based on the results in \cite{3}, \cite{4}.

\section{Geodesic flows on homogeneous spaces with central metrics}

Let $G$ be a connected real $n$-dimensional Lie group, ${\mathfrak g}$ be its Lie
algebra, $H$ be an $(n-m)$-dimensional subgroup of the group $G$, ${\mathfrak h}$ be the
Lie algebra of the group $H$, $M = G/H$ be the homogeneous right $m$-dimensional
$G$-space, and $\pi:\ G \rightarrow M$ be the canonical projection of the group $G$ onto
the space $M$ of right cosets.

Let ${\mathbf G}(\cdot,\cdot)$ be a nondegenerate quadratic form on the cotangent plane
at the unity of the group ${\mathfrak g}^*\cong T_e^*G$. The nondegenerate form $\mathbf
G$ determines a metric on the group $G$ at the unity. Acting on the form $\mathbf G$ by
right and left shifts, we obtain right-invariant (${\mathbf G}^R_g=R_g^* {\mathbf G}$)
and left-invariant (${\mathbf G}^L_g=L_g^* {\mathbf G}$) metrics on the group at an
arbitrary point $g\in G$,
\begin{equation}\label{eq1}
\mathbf{G}^R_g(\sigma,\tau)=\mathbf{G}(R^*_{g^{-1}}\,\sigma,R^*_{g^{-1}}\,\tau),\quad
\mathbf{G}^L_g(\sigma,\tau)=\mathbf{G}(L^*_{g^{-1}}\,\sigma,L^*_{g^{-1}}\,\tau),
\end{equation}
where $\sigma,\tau \in T^*_g G$. We project the right-invariant metric on $M$:
$\mathbf{G}^R_x=\pi_*\mathbf{G}^R_g$, $x=\pi(g)$:
\begin{equation}\label{eq2}
\mathbf{G}^R_x(\sigma, \tau)\equiv \mathbf{G}^R_g(\pi^* \sigma,\pi^*\tau),\quad
\sigma,\tau \in T^*_x M.
\end{equation}
Because the left shift $g \rightarrow hg$ by an arbitrary element $h\in H$ takes each
point $x \in M$ to itself, we can show that definition (\ref{eq2}) of the quadratic form
$\mathbf{G}^R_x$ is correct only if the form $\mathbf{G}$ is $\mathrm{Ad}_H^*$-invariant,
\begin{equation}\label{eq3}
\mathbf{G}(\mathrm{Ad}^*_h\cdot,\mathrm{Ad}^*_h\cdot)|_{\,{\mathfrak
h}^{\perp}}=\mathbf{G}(\cdot,\cdot)|_{\,{\mathfrak h}^{\perp}}.
\end{equation}

Because the group $G$ acts on $M$ by right shifts and the metric $\mathbf{G}^R_g$ is
right-invariant, the quadratic form $\mathbf{G}^R_x$ on $T^*_x M$ is $G$-invariant for
every $\mathrm{Ad}_H^*$-invariant form $\mathbf{G}$. The obvious equality
$\pi_*(\mathfrak{h})=0$ implies that at the point $x_0=\pi(H) \in M$, the space
$T^*_{x_0}M$ is naturally identified with the space $\mathfrak{h}^{\perp}$,
$$
T^*_{x_0}M\cong \mathfrak{h}^{\perp}=\{\lambda \in
\mathfrak{g}^*\,|\,\langle\lambda,\mathfrak{h}\rangle=0\} \subset \mathfrak{g}^*.
$$
The form $\mathbf{G}^R_{x_0}$ at the point $x_0$ is therefore nondegenerate if
\begin{equation}\label{eq4}
\mathrm{rank}\mathbf{G}(\cdot,\cdot)|_{\,\mathfrak{h}^{\perp}}=\dim M.
\end{equation}

By the $G$-invariance of the quadratic form $\mathbf{G}^R_x$, condition (\ref{eq4}) is
not only necessary but also sufficient for the nondegeneracy of this form at an arbitrary
point $x \in M$. The nondegenerate quadratic form $\mathbf{G}$ on $\mathfrak{g}^*$
satisfying conditions (\ref{eq3}) and (\ref{eq4}) thus determines the $G$-invariant
Riemannian metric $\mathbf{G}^R_x$ on $M$ by formula (\ref{eq2}). We note that the
signature of this Riemannian metric is determined by the signature of the quadratic form
$\mathbf{G}|_{\,\mathfrak{h}^{\perp}}$.

Analogously, we define the quadratic form $\mathbf{G}^L_x$ on $T^*_x M$ by projecting the
left-invariant metric $\mathbf{G}^L_x = \pi_* \mathbf{G}^L_g$,

\begin{equation}\label{eq5}
\mathbf{G}^L_x(\sigma, \tau)\equiv \mathbf{G}^L_g(\pi^* \sigma,\pi^*\tau),\quad
\sigma,\tau \in T^*_x M.
\end{equation}
By the left-invariance of the metric $\mathbf{G}^L_g$, the quadratic form
$\mathbf{G}^L_x$ is well defined by (\ref{eq5}) for any choice of the form $\mathbf{G}$.
Condition (\ref{eq4}) is the condition for the nondegeneracy of the form $\mathbf{G}^L_x$
in a neighborhood of the point $x_0$. It can be shown that the form $\mathbf{G}^L_x$ is
nondegenerate at all points $x \in M$ if
\begin{equation}\label{eq6}
\mathrm{rank}
\mathbf{G}(\mathrm{Ad}_g^*\cdot,\mathrm{Ad}_g^*\cdot)|_{\,\mathfrak{h}^{\perp}}=\dim M.
\end{equation}

The nondegenerate quadratic form $\mathbf{G}^L_x$ (\ref{eq5}) obtained by projecting the
left-invariant metric $\mathbf{G}^L_g$ (\ref{eq1}) on the right coset space $M = G/H$
determines a Riemannian metric on $M$, which we call the central metric. Because
condition (\ref{eq3}) is not necessarily satisfied in this case, central metrics form a
wider class of metrics connected with the Lie group on a homogeneous space than
$G$-invariant metrics (\ref{eq2}).

Let $\{e_A\}$ be a basis of the algebra $\mathfrak{g}$, and let $\{e^A\}$ be the dual
basis of $\mathfrak{g}^*$, $\langle e^A,e_B\rangle = \delta^A_B$. The linear space
$\mathfrak{g}$ can be viewed as the dual space of $\mathfrak{g}^*$. The form $\mathbf{G}\
$ and the left-invariant metric $\mathbf{G}^L_g$ on the group $G$ are then written as
$$
\mathbf{G}=\mathbf{G}^{AB} e_A \otimes e_B,\quad \mathbf{G}(e^A,e^B)=\mathbf{G}^{CD}
e_C(e^A)\otimes e_D(e^B)=\mathbf{G}^{AB},
$$
\begin{equation}\label{eq7}
\mathbf{G}^L_g=L^*_g \mathbf{G}=\mathbf{G}^{AB} [(L_g)_* e_A] \otimes [(L_g)_* e_B] =
\mathbf{G}^{AB} \xi_A(g) \otimes \xi_B(g),
\end{equation}
where $\xi_A(g)$ are the left-invariant vector fields on the group $G$. Let $\{z^i\},\ i
= 1,\dots,n$, be the coordinates of an element $g \in G$. In these coordinates, the
components of the left-invariant metric are given by
\begin{equation}\label{eq8}
g^{ij}(z)=\mathbf{G}^L_g(dz^i,dz^j)=\mathbf{G}^{AB} \xi_A^i(z)\xi_B^j(z).
\end{equation}

Any element of the group can be represented as $g=hs(x)$, where $h \in H$ and $s(x)$ is a
smooth local section $s:\ G \rightarrow M$ of the principal bundle $(G,M,\pi,H)$, $\pi
\circ s = \mathrm{id}$. Let $\{y^{\bar a}\}$, $\bar{a}=1,\dots,n-m$, and $\{x^a\}$, $a=1,
\dots,m$, be the respective local coordinates on the fiber $H$ and on the base $M$. In
these coordinates, we have
\begin{equation}\label{eq9}
\xi_A(g)=X_A^a(x)\partial_{x^a}+\xi_A^{\bar{a}}(y,x)\partial_{y^{\bar{a}}},\quad \pi_*
\xi_A(g)=X_A^a(x)\partial_{x^a}.
\end{equation}
Obviously, the vector fields $X_A(x, \partial_x)=X^a_A(x)\partial_{x^a}$ form the Lie
algebra $\mathfrak{g}$, $[X_A,X_B]=C^C_{AB}X_C$, and are the generators of the
transformation group $G$ acting on $M$ (here, $C^C_{AB}$ are the structure constants of
the algebra $\mathfrak{g}$ in the basis $\{e_A\}$). In the local coordinates, the
components of the central metric are given by
\begin{equation}\label{eq10}
g^{ab}(x)=\mathbf{G}^{AB}X_A^a(x)X_B^b(x).
\end{equation}

We show how the geodesics for the respective metrics (\ref{eq8}) and (\ref{eq10}) on $G$
and $M$ are related (we assume that the connections are compatible with the metrics).

The manifold $T^* G$ is endowed by the natural simplectic structure
\begin{equation}\label{eq11}
p=p_idz^i,\quad \omega=dp=dp_i \wedge dz^i.
\end{equation}
By relations (\ref{eq11}), the geodesic equation
$$
\frac{d^2 z^i}{dt^2}+ \Gamma^i_{jk} \frac{dz^j}{dt} \frac{dz^k}{dt}=0
$$
can be represented in the Hamiltonian form
\begin{equation}\label{eq12}
\frac{dz^i}{dt}=\{\widetilde{H},z^i\},\quad\frac{dp_i}{dt}=\{\widetilde{H},p_i\}
\end{equation}
with the Hamiltonian $\widetilde{H}(z, p)=(1/2)g^{ij}(z)p_ip_j$.

The left-invariant vector fields $\xi_A(g)$ form a basis of the tangent space $T_g G$ at
each point $g \in G$. Analogously, the left-invariant 1-forms $\omega^A(g)=\omega^A_i(z)
dz^i$ form a basis of the cotangent space $T^*_g G$, $\xi^i_A\omega^A_j=\delta^i_j$,
$\xi^i_A \omega^B_i=\delta^B_A$, and every vector $T^i$ or covector $T_i$ is decomposed
in this basis: $T^i=T^A\xi^i_A$, $T_i=T_A\omega^A_i$. Obviously, a tensor of any order
can be decomposed in this "vierbein" basis; in particular, the components of the matrix
$\mathbf{G}^{AB}$ are the "vierbein" components of metric (\ref{eq8}). In the "vierbein"
components $P_A=\xi^i_A p_i$, geodesic equations (\ref{eq12}) assume the triangular form
\begin{equation}\label{eq13}
\frac{dP_A}{dt}=\{H,P_A\}^{\mathrm{Lie}},
\end{equation}
\begin{equation}\label{eq14}
\frac{dz^i}{dt}=\mathbf{G}^{AB}P_A(t)\xi_B^i(z),
\end{equation}
where $H(P)=(1/2)\mathbf{G}^{AB}P_AP_B$ and $\{\cdot,\cdot\}^{\mathrm{Lie}}$ is the
Poisson–Lie bracket on $\mathfrak{g}^*$,
$$
\{\varphi,\psi\}^{\mathrm{Lie}}=\langle P,[\nabla\varphi,\nabla\psi]\rangle=C_{AB}^C P_C
\frac{\partial \varphi(P)}{\partial P_A} \frac{\partial \psi(P)}{\partial P_B},
$$
$$
P=P_A e^A \in \mathfrak{g}^*,\quad \varphi,\psi \in C^{\infty}(\mathfrak{g}^*).
$$

The mapping $\mu: T^*G \rightarrow \mathfrak{g}^*$ defined by the relation $\mu(z,p)=P_A
e^A$ with $P_A=\xi^i_A(z)p_i\ $ is called the moment mapping.

Integrating the geodesic flow for the left-invariant metric is thus reduced to
integrating a Hamiltonian system on the coalgebra $\mathfrak{g}^*$ given by (\ref{eq13})
and then finding the integral trajectory of the left-invariant nonautonomous vector field
\begin{equation}\label{eq15}
\xi(t,g)=\mathbf{G}^{AB} P_A(t)\xi_B(g).
\end{equation}

We introduce the simplectic form $\omega$ on $T^*M$, $p=p_adx^a$, $\omega=dp$, and
represent the geodesic equations on the Riemannian space $M$ with central metric
(\ref{eq10}) in the Hamiltonian form
\begin{equation}\label{eq16}
\frac{dx^a}{dt}=\{H,x^a\},\quad \frac{dp_a}{dt}=\{H,p_a\}
\end{equation}
where
$$
H(x,p)=\frac12\, \mathbf{G}^{AB} X_A (x,p) X_B(x,p),\quad X_A(x,p)\equiv X_A^a(x)p_a.
$$
We introduce the moment mapping $\mu$: $T^*M \rightarrow \mathfrak{g}^*$ in a standard
way, $\mu(x, p)=P_Ae^A$, where $P_A=X_A(x,p)$, and represent Hamiltonian system
(\ref{eq16}) in the form of equations (\ref{eq13}) and the equations
\begin{equation}\label{eq17}
\frac{dx^a}{dt}=\mathbf{G}^{AB}P_A(t)X_B^a(x).
\end{equation}
We see that integrating the geodesic flow with the central metric is reduced to
integrating a Hamiltonian system on coalgebra (\ref{eq13}) and finding integral
trajectories of the nonautonomous vector field
$X(t,x)=\mathbf{G}^{AB}P_A(t)X^a_B(x)\partial_{x^a}$, which is the projection on $M$ of
the left-invariant vector field $\xi(t,g)$ given by (\ref{eq15}). Moreover, it follows
from formula (\ref{eq9}) that system of equations (\ref{eq14}) consists of system
(\ref{eq17}) and the additional system
$$
\frac{dy^{\bar{a}}}{dt}=\mathbf{G}^{AB}P_A(t)\xi_B^{\bar{a}}(y,x).
$$
Geodesic equations (\ref{eq16}) for central metric (\ref{eq10}) are thus a part of
geodesic equations (\ref{eq12}) for left-invariant metric (\ref{eq8}).

We note that in the case of the left-invariant metric, the coordinates $P_A$ of the
covector $P=P_Ae^A \in \mathfrak{g}^*$ of system (\ref{eq13}) at the initial instant can
be arbitrary, while in the case of the central metric, the initial values $P_A(0)$ (as
well as the whole trajectory $P_A(t)$) lie on the invariant surface
$\mathrm{Ad}_G^*\,\mathfrak{h}^{\perp}\subset \mathfrak{g}^*$, where
$\mathrm{Ad}_G^*\,\mathfrak{h}^{\perp}$ is the orbit of the coadjoint representation of
the group $G$ of the space $\mathfrak{h}^{\perp}$. In other words, the image of the
simplectic manifold $T^*M$ under the moment mapping $\mu(T^*M)$ is not the entire space
$\mathfrak{g}^*$ but its invariant subspace $\mathrm{Ad}_G^*\,\mathfrak{h}^{\perp}$.

Indeed, we first set $x(0)=x_0=\pi(H)$. Then by definition of the isotropy subalgebra, we
have $P^0_{\bar{a}}=X^b_{\bar{a}}(x_0)p^0_b=X_{\bar{a}}(x_0,p^0)=0$, which is equivalent
to the equality $\langle P^0,e_{\bar{a}} \rangle = 0$ or $\langle
P^0,\mathfrak{h}\rangle=0$, i.e., $P^0 \in \mathfrak{h}^{\perp}$. Now let the initial
point $x=x(0)$ not coincide with the point $x_0$. Obviously, there is an element $g$ of
the group $G$ such that $x=x_0g$. It can be proved (see, e.g., \cite{4}) that the
equality $X_A(x,p)=(\mathrm{Ad}_g)^B_A X_B(x_0,p^0)$ holds, where $p=(R_g)^*p^0$. We thus
have $P_A(0)=(\mathrm{Ad}^*g^{-1} P^0)_A$, i.e., $P(0)=\mathrm{Ad}^*_{g^{-1}} P^0 \in
\mathrm{Ad}_G^*\,\mathfrak{h}^{\perp}$. (in other words, the moment mapping is an
equivariant mapping).

Above, we have considered the connection between geodesic flows with left-invariant and
central metrics. In the consideration of integrability questions below, it is convenient
to assume that the left-invariant metric is central and the group $G$ in this case is a
homogeneous right $G$-space with the trivial stationary subgroup.

\section{The construction of canonical transformation}

The algebra $\mathfrak{g}$ of the operators $X_A(x,\partial_x)=X_A^a(x)\partial_{x^a}$
generally allows for the algebra of invariant differential and pseudodifferential
operators $L(x, \partial_x)$:
$$
[L(x,\partial_x),X_A(x,\partial_x)]=0.
$$
For commutative spaces, the algebra of invariant operators is the center of the
enveloping field $E(\mathfrak{g})$ and consists of the Casimir operators
$K(X(x,\partial_x))$. Analogously, the functions $X_A(x,p)=X^a_A(x)p_a$ on $T^*M$ form
the same Lie algebra $\mathfrak{g}$ with respect to the Poisson bracket and allow for the
algebra of functions $L(x,p)$ invariant under Poisson transformations generated by the
algebra $\mathfrak{g}$: $\{L(x,p),X_A(x,p)\}=0$. We choose a basis of independent
invariant functions $L_{\mu}(x,p)$. It is obvious that the Poisson bracket of the
functions $L_{\mu}(x,p)$ is functionally expressed in terms of this set,
\begin{equation}\label{eq18}
\{l_{\mu},L_{\nu}\}=\Omega_{\mu\nu}(L).
\end{equation}
The space ${\cal F}$ with the basis $\{L_{\mu}(x,p)\}$ and nonlinear commutation
relations (\ref{eq18}) is called the functional algebra or $\cal F$-algebra (quadratic
algebra, cubic algebra, and so on).

It is clear that the functions $\{L_{\mu}(x,p)\}$ are independent integrals of motion of
Hamiltonian system (\ref{eq16}), and in view of the arbitrariness of the form
$\mathbf{G}$, this system has no other symmetries. Passing to triangular form
(\ref{eq13}) and (\ref{eq17}) also might not lead to the solution of the posed problem.
Not only solving Eqs. (\ref{eq13}) but also integrating system (\ref{eq17}) can present
difficulties because it is nonautonomous. For example, in the left-invariant metric case,
system (\ref{eq14}) allows for the symmetry algebra $\mathfrak{g}$ of right-invariant
vector fields for arbitrary functions $P_A(t)$. But according to the Lie theorem, an
$n$-dimensional system of ordinary differential equations is integrable if its symmetry
algebra is solvable, which is generally not the case. Below, we show that the
integrability of the system on coalgebra (\ref{eq13}) is sufficient for solving geodesic
flow equations (\ref{eq16}).

Hamiltonian systems are integrated as follows. The system is first restricted to the
level surface of the integrals of motion. The reduced system has a commutative symmetry
group generated by Hamiltonian flows of some integrals of motion whose vector fields lie
on the level surface of the integrals of motion. As mentioned in the introduction, the
level surface is ill-defined in some cases, and the traditional method is inapplicable.
Below, we describe an alternative way to integrate Hamiltonian systems based on
constructing a special canonical transformation. We construct this transformation using
local coordinates, and we therefore do not discuss global questions.

The maximal dimension of the orbits of the coadjoint representation (K-orbits) is equal
to $n-r$, where the number $r=\mathrm{ind\,} \mathfrak{g}$ is called the index of the
algebra $\mathfrak{g}$. In general, the invariant space
$\mathrm{Ad}_G^*\,\mathfrak{h}^{\perp}$ consists of singular orbits. Let the maximal
dimension of K-orbits belonging to the space $\mathrm{Ad}_G^*\,\mathfrak{h}^{\perp}$ be
equal to $n-r-2s_M$. The number $s_M$ is called the degree of degeneracy of the
homogeneous space $M$ and is determined by the structure constants of the algebra
$\mathfrak{g}$ and subalgebra $\mathfrak{h}$ by the formula \cite{4}
\begin{equation}\label{eq19}
s_M=\frac12\, (\dim \mathfrak{g}^{\lambda}-\mathrm{ind\,}\mathfrak{g}),
\end{equation}
where $\lambda$ is a general element of the space $\mathfrak{h}^{\perp}$,
$\mathfrak{g}^{\lambda}$ is the annihilator of the covector $\lambda$.

In general, the functions $X_A(x,p)$ are not functionally independent, and there are
functional relations $\Gamma(X(x,p))=0$ between them, which are called identities. The
number of independent identities $i_M$ is called the index of a homogeneous space. As
shown in \cite{4}, the index of a homogeneous space satisfies the relation
$i_M=\mathrm{codim\,}\mathrm{Ad}_G^*\,\mathfrak{h}^{\perp}$ and can be found from the
structure constants of the algebra $\mathfrak{g}$ and its subalgebra $\mathfrak{h}$,
\begin{equation}\label{eq20}
i_M=\dim \mathfrak{h}^{\lambda},\quad \mathfrak{h}^{\lambda}= \mathfrak{h} \cap
\mathfrak{g}^{\lambda}.
\end{equation}

We introduce the dual space ${\cal F}^*=\{a_{\mu}\}$ of $\cal F$ with the generators
$a_{\mu}$, and we define the Poisson bracket
\begin{equation}\label{eq21}
\{\varphi,\psi\}^{\cal F}(a)=\Omega_{\mu\nu}(a)\frac{\partial \varphi(a)}{\partial
a_{\mu}} \frac{\partial \psi(a)}{\partial a_{\nu}}
\end{equation}
in the space of smooth functions on ${\cal F}^*$. Then $C^{\infty}({\cal F})$ becomes a
Poisson algebra. We also introduce the moment mapping $\widetilde{\mu}$: $T^*M
\rightarrow {\cal F}$ by the relation $L_{\mu}(x,p)=a_{\mu}$. The moment mappings $\mu$:
$T^*M \rightarrow \mathrm{Ad}_G^*\,\mathfrak{h}^{\perp}$ and $\widetilde{\mu}$: $T^*M
\rightarrow {\cal F}$ determine a bifibration. As shown in \cite{5}, the simplectic
sheets $\Omega \subset \mathrm{Ad}_G^*\,\mathfrak{h}^{\perp}$ and $\widetilde{\Omega}
\subset {\cal F}^*$ are in one-to-one correspondence:
$\widetilde{\Omega}=\widetilde{\mu}(\mu^{-1}(\Omega))$. This means that the number of
nontrivial Casimir functions $K_m(P)$ on $\mathrm{Ad}_G^*\,\mathfrak{h}^{\perp}$
coincides with the number of Casimir functions $Z_m(a)$ on ${\cal F}^*$ and they can be
chosen compatible with each other:
\begin{equation}\label{eq22}
Z_m \circ \widetilde{\mu}=K_m \circ \mu \Longleftrightarrow Z_m(L(x,p))=K_m(X(x,p)).
\end{equation}
By analogy with Lie algebras, we call the number of nontrivial Casimir functions on
${\cal F}^*$ the index ($\mathrm{ind}\, {\cal F}$) of the $\cal F$-algebra. The
compatible simplectic sheets $\Omega_{\kappa}$ and $\widetilde{\Omega}_{\kappa}$ indexed
by the ($\mathrm{ind}\, {\cal F}$)-dimensional parameter $\kappa$ are defined by
\begin{equation}\label{eq23}
\Omega_{\kappa}=\{P\in \mathrm{Ad}_G^*\,\mathfrak{h}^{\perp}\,|\,K_m(P)=\kappa_m\},\quad
\widetilde{\Omega}_{\kappa}=\{a\in {\cal F}^*\,|\,Z_m(a)=\kappa_m\}.
\end{equation}

The dimension and index of the $\cal F$-algebra are given by the formulas \cite{4}
\begin{equation}\label{eq24}
\dim {\cal F}=i_M+2 \dim M - \dim \mathfrak{g},\quad \mathrm{ind\,}{\cal
F}=\mathrm{ind\,}\mathfrak{g}+2s_M-i_M.
\end{equation}
The dimension of the simplectic sheet $\widetilde{\Omega}$ is $\dim \widetilde{\Omega}=
\dim {\cal F} - \mathrm{ind\,} {\cal F} = 2d(M)$. The nonnegative integer $d(M)$ is
called the defect of the homogeneous space $M$. For spaces with zero defect, the algebra
of invariant operators is commutative and belongs to the center of the enveloping field
\cite{4}. Spaces with zero defect are therefore commutative spaces. The class of
commutative spaces includes all symmetric and weakly symmetric spaces \cite{6}. In
\cite{7}, commutative spaces were studied by geometric methods. In this connection, we
note that to establish the commutativity of a space, it suffices to find its defect
\begin{equation}\label{eq25}
d(M)=\frac12\,\dim \mathfrak{g}/\mathfrak{g}^{\lambda}-\dim
\mathfrak{h}/\mathfrak{h}^{\lambda}.
\end{equation}

Let the ($\mathrm{ind\,}{\cal F}$)-dimensional real parameter $j$ index the orbits of
maximal dimension in $\mathrm{Ad}_G^*\,\mathfrak{h}^{\perp}$, and let $\lambda(j)$ be a
representative of the orbit ${\cal O}_{\lambda(j)} \subset
\mathrm{Ad}_G^*\,\mathfrak{h}^{\perp}$. On the orbit ${\cal O}_{\lambda(j)}$, we
introduce the Darboux coordinates ($\pi_{\alpha},q^{\alpha}$) in which the Kirillov
simplectic form \cite{1} has the canonical form
$$
\omega_{\lambda}=\sum \limits_{\alpha=1}^{(n-r)/2-s_M} d\pi_{\alpha} \wedge dq^{\alpha}.
$$
In the overwhelming majority of cases, the covector $\lambda \in \mathfrak{g}^*$ has a
normal polarization $\mathfrak{p} \subset \mathfrak{g}^{\mathbb{C}}$:
$$
\dim \mathfrak{p}=\dim \mathfrak{g} - \frac12\, \dim{\cal O}_{\lambda},\quad
\langle\lambda,[\mathfrak{p},\mathfrak{p}]\rangle=0.
$$
In this case, the transition to the canonical coordinates is determined by the expression
\cite{3}
\begin{equation}\label{eq26}
P_A=P_A(q,\pi,j)=P^{\alpha}_A(q)\pi_{\alpha}+\chi_A(q,\lambda(j)),
\end{equation}
and we have
\begin{equation}\label{eq27}
K_m(P(q,\pi,j))=\kappa_m(j),\quad \det \left \| \frac{\partial \kappa_m(j)}{\partial j_k}
\right \|\neq 0.
\end{equation}

Analogously, we pass to the canonical coordinates $(u, v)$ on the simplectic sheet
$\widetilde{\Omega}_{\kappa(j)}$ given by (\ref{eq23}):
\begin{equation}\label{eq28}
a_{\mu}=a_{\mu}(u,v,j);\quad Z_m(a(u,v,j))=\kappa_m(j).
\end{equation}

We extend the simplectic space $\Omega \times \widetilde{\Omega}$ with the form $d\pi
\wedge dq + dv \wedge du$ up to a simplectic space $K$ homeomorphic to the space $T^*M$
by adding $\mathrm{ind\,}{\cal F}$ pairs of canonically conjugate quantities $(\tau, j)$
and defining the simplectic form
\begin{equation}\label{eq29}
\widetilde{\omega}=\sum \limits_{\alpha=1}^{(n-r)/2-s_M} d\pi_{\alpha} \wedge
dq^{\alpha}+\sum \limits_{\bar{\alpha}=1}^{d(M)} dv_{\bar{\alpha}} \wedge
du^{\bar{\alpha}}+\sum \limits_{m=1}^{\mathrm{ind\,}{\cal F}} dj_m \wedge d\tau^m
\end{equation}
on $K$. We further define a (locally) one-to-one transition from the coordinates $(x, p)$
on $T^*M$ to the coordinates $(q,\pi,u,v,j,\tau)$ on $K$ such that
$\omega=\widetilde{\omega}$.

The moment mappings $\mu$ and $\widetilde{\mu}$ are Poisson mappings $T^*M \rightarrow
K$,
$$
\{\varphi \circ \mu,\psi \circ \mu\}=\{\varphi,\psi\}^{\mathrm{Lie}}\circ \mu,\quad
\{f\circ \widetilde{\mu},h \circ \widetilde{\mu}\}=\{f,h\}^{\cal F}\circ \widetilde{\mu},
$$
where $\varphi,\psi \in C^{\infty}(\mathfrak{g}^*)$ and $f,h \in C^{\infty} ({\cal
F}^*)$. In local coordinates, they have the form
\begin{equation}\label{eq30}
X_A(x,p)=P_A(q,\pi,j),\quad L_{\mu}(x,p)=a_{\mu}(u,v,j).
\end{equation}
Relations (\ref{eq30}) implicitly define the variables $(q,p,u,v,j)$ as functions of the
variables $(x, p)$. In particular, the equalities $K_m(X(x,p))=\kappa_m(j)$ implicitly
define the functions $j_m = j_m(x,p)$.

We also define the mapping $T$ : $T^*M \rightarrow K$ by the relation
\begin{equation}\label{eq31}
T^m(x,p)=\tau^m.
\end{equation}
Obviously, the mapping $\Lambda=(\mu,\widetilde{\mu},T): T^*M \rightarrow K$ is
simplectic, i.e., takes the form $\omega$ to the form $\widetilde{\omega}$, if and only
if the functions $T^m(x,p)$ satisfy the equations
\begin{equation}\label{eq32}
\{T^m,j_k\}=\delta_k^m,\quad
\{T^m,q^{\alpha}\}=\{T^m,\pi_{\alpha}\}=\{T^m,u^{\bar{\alpha}}\}=\{T^m,v_{\bar{\alpha}}\}=0.
\end{equation}
Here, we assume that the variables $q, \pi, u, v$, and $j$ are functions of $x$ and $p$
determined by relations (\ref{eq30}).

Because the forms $\omega$ and $\widetilde{\omega}$ are nondegenerate and the dimension
of the space $T^*M$ is equal to that of the space $K$, the mapping $\Lambda$ defined by
relations (\ref{eq30}) and (\ref{eq31}) is a local simplectic one-to-one mapping and
therefore realizes the required canonical transformation.

We note that in the case of a commutative group $G$, the simplectic sheets $\Omega$ and
$\widetilde{\Omega}$ are zero-dimensional and there are no coordinates $q, \pi, u$, and
$v$. In this case, the canonical transformation consists in passing to "action–angle"
variables, where $j$ are the "action" variables and $\tau$ are "angle" variables.

\section{Integration of a geodesic flow with G-invariant metrics}

Because for a $G$-invariant metric, the Hamiltonian of the geodesic flow is a
$G$-invariant function under the induced action of the group on $T^*M$, it is a function
of the basic invariants $L_{\mu}(x, p)$, i.e., $H=H(L(x,p))$. The same holds for an
arbitrary $G$-invariant function $H(x,p)$ on $T^*M$. For this reason, we study the
integrability of system (\ref{eq16}), where the Hamiltonian $H$ is an arbitrary
$G$-invariant function on $T^*M$.

After canonical transformation (\ref{eq30}), (\ref{eq31}), the Hamiltonian $H(L(x,p))$
passes to the Hamiltonian $\widetilde{H}(u,v,j)=H(a(u,v,j))$ , and Eqs. (\ref{eq16})
become
\begin{equation}\label{eq33}
\frac{du^{\bar{\alpha}}}{dt}=\frac{\partial \widetilde{H}(u,v,j)}{\partial
v_{\bar{\alpha}}},\quad
\frac{dv_{\bar{\alpha}}}{dt}=-\frac{\partial
\widetilde{H}(u,v,j)}{\partial u^{\bar{\alpha}}},
\end{equation}
\begin{equation}\label{eq34}
\frac{dq}{dt}=\frac{d\pi}{dt}=\frac{dj}{dt}=0,\quad \frac{d\tau^m}{dt}=\frac{\partial
\widetilde{H}(u,v,j)}{\partial j_m}.
\end{equation}

System of equations (\ref{eq34}) is integrated elementarily if the solution of system
(\ref{eq33}) is known. The integrability of the initial G-invariant Hamiltonian system is
thus equivalent to the integrability of system (\ref{eq33}).

It is clear that Hamiltonian system (\ref{eq33}) is the result of the reduction of
initial Hamiltonian system (\ref{eq16}) on the $2d(M)$-dimensional simplectic sheet
$\widetilde{\Omega}_{\kappa(j)}$. For $d(M)=0$ (which corresponds to the commutative
space case), there are no variables $u$ and $v$ and no system (\ref{eq33}). Hamiltonian
system (\ref{eq33}) is also integrable for $d(M)=1$. In this case, using the Hamiltonian
$H$ as an integral of motion, we can easily obtain solutions in quadratures. It is
obvious that system (\ref{eq33}), in general, is not integrable for $d(M)>1$.

We note that finding the algebra of invariant functions and constructing the canonical
variables on the simplectic sheets $\Omega$ and $\widetilde{\Omega}$ and the functions
$T^m(x,p)$ is reduced to quadratures. We thus obtain the following statement.

\begin{theorem}
An arbitrary G-invariant Hamiltonian system on $T^*M$ reduces to an autonomous
$2d(M)$-dimensional Hamiltonian system. In particular, it is integrable in quadratures if
and only if $d(M) < 2$.
\end{theorem}

In \cite{8}, \cite{9}, a criterion for the integrability of arbitrary $G$-invariant
Hamiltonian systems on $T^*M$ in the class of Noether integrals is given. In our
notation, this criterion has the form
$$
\frac12\,\dim \mathfrak{g}/\mathfrak{g}^{\lambda}+\dim
\mathfrak{g}^{\lambda}/\mathfrak{h}^{\lambda}=\dim \mathfrak{g}/\mathfrak{h}.
$$
It follows from formula (\ref{eq25}) that this condition is the zero-defect condition,
$d(M)=0$. But there is no contradiction with Theorem 1, because for $d(M)=1$, the
Hamiltonian $H$ of system (\ref{eq16}) does not belong to the class of Noether integrals.
We recall that the class of Noether integrals consists of the functions on $T^*M$ of the
form $\varphi\circ\mu$, $\varphi \in C^{\infty}(\mathfrak{g}^*)$. As a consequence, we
obtain the following criterion for the commutativity of a homogeneous space.

\begin{corollary}
An arbitrary $G$-invariant Hamiltonian system on $T^*M$ is integrable in the class of
Noether integrals if and only if the space $M$ is commutative.
\end{corollary}

As an example, we consider the unsolvable five-dimensional Lie group $G$ whose algebra
$\mathfrak{g}$ has the following nonzero commutation relations:
$$
[e_1,e_4]=-e_1,\quad [e_1,e_5]=e_2,\quad [e_2,e_3]=e_1,\quad [e_2,e_4]=e_2,
$$
$$
[e_3,e_4]=-2e_3,\quad [e_3,e_5]=e_4,\quad [e_4,e_5]=-2e_5.
$$
It is easy to verify that $r=\mathrm{ind\,}\mathfrak{g}=1$.

Let $M=G/H$ be a four-dimensional homogeneous space with a one-dimensional isotropy
subgroup $H$ whose one-dimensional Lie algebra $\mathfrak{h}$ is generated by the vector
$e_5$. By (\ref{eq19}), (\ref{eq20}), (\ref{eq24}), and (\ref{eq25}), we have $s_M=0$,
$i_M=0$, $\dim {\cal F}=3$, $\mathrm{ind\,}{\cal F}=1$, and $d(M)=1$. Therefore, the
space $M$ is noncommutative, the algebra of invariant functions has three generators and
allows for one independent Casimir function, and an arbitrary $G$-invariant Hamiltonian
system on $T^*M$ is integrable in quadratures but not in the class of Noether integrals.

We introduce the local coordinates $x$ on the group $G$ by the relation
$g=e^{x_5e_5}\dots e^{x_1e_1}$. In these coordinates, the functions $X(x,p)$ and $L(x,p)$
are given by
$$
X_1=p_1,\quad X_2=p_2,\quad X_3=x_2p_1+p_3,
$$
$$
X_4=-x_1p_1 + x_2p_2 - 2x_3p_3 + p_4,\quad X_5=x_1p_2 - x^2_3 p_3 + x_3p_4,
$$
$$
L_1=-e^{x_4}(x_3p_1 + p_2),\quad L_2=-p_4,\quad L_3=e^{-x_4}(p_1p_4 - x_3p_1p_3 -
p_2p_3).
$$
An algorithm for finding the algebra of invariant operators on homogeneous spaces is
given in \cite{4}, where, in particular, the example of this $G$-space is analyzed.

We consider the Hamiltonian system with the most general Hamiltonian corresponding to the
$G$-invariant metric,
$$
H= H(L(x, p))=\frac12\, c_1L^2_1(x,p)+c_2L^2_2(x,p)+c_3L_1(x,p)L_2(x,p)+c_4L_3(x,p),
$$
where $c_i$ are constants. We note that the metric under consideration is
non-St\"{a}ckel, i.e., the corresponding Hamilton–Jacobi equation cannot be integrated
using the variable separation method.

The canonical coordinates corresponding to the polarization
$\mathfrak{p}=\{e_1,e_2,e_5\}$ on the nondegenerate K-orbit passing through the covector
$\lambda=(1,0,0,0,j)$ are given by
$$
P_1=q_1,\quad P_2=-q_2,\quad P_3=q_1\pi_2,\quad P_4=-q_2\pi_2+q_1\pi_1,\quad
P_5=q_2\pi_1+\frac{j}{q_1^2}.
$$
In this case, there is only one Casimir function $K=P_1P_2P_4+P^2_1 P_5 -P^2_2 P_3=j$.
The basis elements of the algebra of invariant functions satisfy the commutation
relations
$$
\{L_1, L_2\}=L_3,\quad \{L_1,L_3\}=0,\quad \{L_2, L_3\}=L_3.
$$
In the space $C^{\infty}({\cal F}^*)$, there is one Casimir function $Z(a)=a_1a_3$, which
satisfies the relation $Z(L(x,p))=K(X(x,p))$. It is easy to find the canonical
coordinates on the simplectic sheet $\widetilde{\Omega}=\{a \in {\cal F}^*\, |\,
a_1a_3=j\}$: $a_1=u$,$a_2=-uv$,and $a_3=j/u$. The function
$T(x,p)=[p^2_1(p_1x_3+p_2)]^{-1}$ is also readily found.

The required canonical transformation $(x,p)\rightarrow (q,\pi,u,v,j,\tau)$ is implicitly
determined by the equations
$$
p_1=q_1,\quad p_2=-q_2,\quad x_2p_1+p_3=q_1\pi_2,
$$
$$
-x_1p_1+x_2p_2-2x_3p_3+p_4=q_1\pi_1-q_2\pi_2,
$$
$$
x_1p_2-x^2_3 p_3 + x_3p_4 = q_2\pi_1 + \frac{j}{q^2_1},\quad -e^{x_4}(x_3p_1 +
p_2)=u,\quad p_4=uv,
$$
$$
e^{-x_4}(p_1p_4 - x_3p_1p_3 - p_2p_3) = \frac{j}{u},\quad p^2_1(p_1x_3+p_2)=\tau^{-1}.
$$
The Hamiltonian takes the form
$\widetilde{H}=H(a(u,v,j)=(c_1u^2+c_2u^2v^2-c_3u^2v+c_4j/u)/2$. The Hamiltonian system
with the Hamiltonian $\widetilde{H}$ and the phase variables $(u,v)$ is easily integrated
in quadratures. In this case, the variables $q$, $\pi$, and $j$ are constants, and the
time-dependence of the variable $\tau$ is obtained by integration,
$$
\tau(t)=\int \frac{\partial\widetilde{H}(u(t),v(t),j)}{\partial j}\, dt = \int \frac{c_4
dt}{u(t)}.
$$

\section{Integration of geodesic flows for central metrics: The case of a wild
Lie algebra}

Geodesic flow equations (\ref{eq16}) for an arbitrary central metric of form (\ref{eq10})
allow for an ${\cal F}$-algebra ${\cal F}=\{L(x,p)\}$ of integrals of motion. After
canonical transformation (\ref{eq30}), (\ref{eq31}), these equations become
\begin{equation}\label{eq35}
\frac{dq^{\alpha}}{dt}=\frac{dq^{\alpha}}{dt}=\frac{\partial
\widetilde{H}(q,\pi,j)}{\partial \pi_{\alpha}},\quad
\frac{d\pi_{\alpha}}{dt}=-\frac{\partial \widetilde{H}(q,\pi,j)}{\partial q^{\alpha}},
\end{equation}
\begin{equation}\label{eq36}
\frac{d\tau^m}{dt}=\frac{\partial \widetilde{H}(q,\pi,j)}{\partial j_m},\quad
\frac{dv_{\bar{\alpha}}}{dt}=\frac{du^{\bar{\alpha}}}{dt}=\frac{dj_m}{dt}=0,
\end{equation}
where $\widetilde{H}(q,\pi,j)=(1/2)\mathbf{G}^{AB}P_A(q,\pi,j)P_B(q,\pi,j)$. System
(\ref{eq36}) is integrated elementarily if the solution of Hamiltonian system
(\ref{eq35}) is known. The number of phase variables $(q,\pi)$ of system (\ref{eq35}) is
equal to $\dim {\cal O}_{\lambda(j)}$, where $\lambda(j) \in \mathfrak{h}^{\perp}$.

\begin{theorem}
The geodesic flow on a homogeneous space with an arbitrary central metric reduces to an
autonomous $\dim {\cal O}_{\lambda}$-dimensional Hamiltonian system and is integrable in
quadratures if and only if
$$
\frac12\, \dim {\cal O}_{\lambda}=\frac{\dim \mathfrak{g}-\mathrm{ind\,}
\mathfrak{g}}{2}-s_M<2.
$$
\end{theorem}

In the case $\dim {\cal O}_{\lambda}=0$, we have $\widetilde{H}=\widetilde{H}(j)$, i.e.,
there are no equations (\ref{eq35}). This case corresponds to the $G$-invariant metric,
where $H(x,p)$ is a Casimir function: $H(x,p)=K\circ\mu$. For $\dim {\cal
O}_{\lambda}=2$, the central metric is not invariant, and this case is more interesting.
We give a nontrivial example illustrating the usefulness of applying canonical
transformation (\ref{eq30}), (\ref{eq31}).

We consider the group $G$ of the matrices of the form
$$
\left(%
\begin{array}{ccc}
  e^{it} &  0 & z \\
  0 &  e^{i\alpha t} & \omega \\
  0 & 0 &  1 \\
\end{array}%
\right),
$$
where $\alpha$ is an irrational number (parameter), $t \in \mathbb{R}^1$, and $z,\omega
\in \mathbb{C}^1$ (this group is given in \cite{1} as an example of a wild Lie group with
a nonsemiseparated K-orbit space). The group $G$ is the semidirect product of the
one-dimensional subgroup $z=\omega=0$ and the four-dimensional normal commutative
subgroup $t=0$; as a topological space, $G$ is homeomorphic to the Euclidean space
$\mathbb{R}^5$. In some basis $\{e_A\}$, the algebra $\mathfrak{g}$ of the group $G$ has
the nonzero commutation relations
$$
[e_1,e_2]=e_3,\quad [e_1,e_3]=-e_2,\quad [e_1,e_4]=\alpha^2 e_5,\quad [e_1,e_5]=-e_4.
$$
In the space $C^{\infty}(\mathfrak{g}^*)$, there are three independent Casimir functions,
i.e., $r=\mathrm{ind\,}\mathfrak{g}=3$. On $\mathfrak{g}^*$, we introduce the "polar"
coordinate system
\begin{equation}
\label{eq37} P_1=\sigma,\quad P_2=\gamma \sin \varphi,\quad P_3=\gamma \cos \varphi,\quad
P_4=\rho \sin \psi,\quad P_5=\frac{\rho}{\alpha} \cos \psi.
\end{equation}
In these coordinates, the Casimir functions are written as $K_1=\gamma$, $K_2=\rho$, and
$K_3=\psi-\alpha\varphi$. Because the variables $\varphi$ and $\psi$ are defined modulo
$2\pi$ and $\alpha$ is an irrational number, the function $K_3$ is infinite-valued, and
at every point of $\mathfrak{g}^*$, its image is a dense subset of the semisegment
$[0,2\pi)$. The level surface for the integral $K_3 \circ \mu$ on $T^*M$ is therefore
undefined. We try to integrate the geodesic flow equations for the central metric on the
four-dimensional homogeneous space.

\begin{floatingfigure}{80mm}
\flushright
\includegraphics[width=75mm]{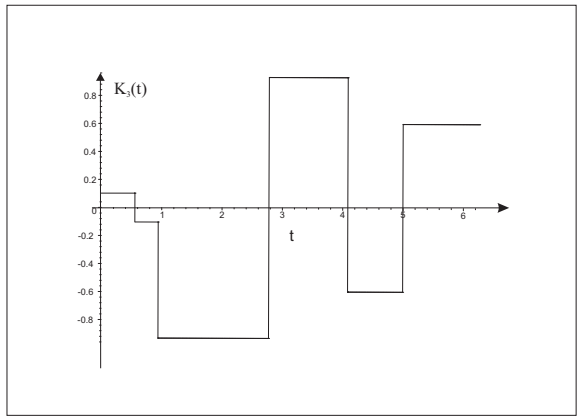}
\caption{function $K_3$} \label{pic:1}
\end{floatingfigure}

For the given motion group, there are three nonequivalent homogeneous spaces with
one-dimensional isotropy subalgebras: $\{e_1\}$, $\{e_2\}$, and $\{e_4\}$. The group $G$
does not act effectively in the spaces with the isotropy subgroups $\{e_2\}$, $\{e_4\}$;
therefore, we consider the homogeneous space $M$ with the isotropy subalgebra $\{e_1\}$
below.

In some coordinates, the functions $X_A(x,p)$ have the form
$$
X_1=x_3p_2-x_2p_3+x_1p_4-\alpha^2 x_4p_1,\quad X_2=p_2,\quad X_3=p_3,\quad X_4=p_4,\quad
X_5=p_1.
$$
As the Hamiltonian of the geodesic flow with central metric (\ref{eq16}), we take the
function
$$
H(x,p)=\frac12\, \mathbf{G}^{AB}X_A(x,p)X_B(x,p),
$$
where $\mathbf{G}^{AB}$ is an arbitrary constant matrix satisfying condition (\ref{eq4}).

By (\ref{eq19}), (\ref{eq20}), (\ref{eq24}), and (\ref{eq25}), we obtain $s_M=0$,
$i_M=0$, $\dim {\cal F}=3$, $\mathrm{ind\,}{\cal F}=3$, $d(M)=0$, and
$\mathrm{Ad}^*_G\,\mathfrak{h}^{\perp}=\mathfrak{g}^*$. Because $\dim{\cal
O}_{\lambda}=2$, Theorem 2 is satisfied, and the Hamiltonian flow is therefore integrable
for an arbitrary central metric on $M$. It is easy to find four commuting integrals for
Hamiltonian system (\ref{eq16}). In addition to the Hamiltonian itself, there are three
Casimir functions $K_m \circ \mu$, $m=1,2,3$. But for the function $K_3 \circ \mu$ (and
any function of this function), it is impossible to define the level surface, and the
standard method for reducing Hamiltonian systems is therefore inapplicable here. We
integrate system (\ref{eq16}) using the canonical transformation.

Using the polarization $\mathfrak{p}=\{e_2,e_3,e_4,e_5\}$ of the functional
$\lambda=(0,0,j_1$, $\alpha j_2 \sin j_3, j_2 \cos j_3)$, we write the canonical
coordinates on the K-orbit ${\cal O}_{\lambda(j)}$,
$$
P_1=\pi,\quad P_2=j_1 \sin q,\quad P_3=j_1 \cos q,\quad P_4=\alpha j_2 \sin(j_3+\alpha
q),
$$
$$
P_5=j_2 \cos(j_3+\alpha q)
$$
(here, we do not consider the special case of orbits passing through the point $j_1=0$ or
$j_2=0$). It is easy to find the connection between the canonical coordinates
$(\pi,q,j_1,j_2,j_3)$ and the "polar" coordinates on $\mathfrak{g}^*$ given by
(\ref{eq37}): $\sigma=\pi$, $\gamma=j_1$, $\varphi=q$, $\rho=\alpha j_2$, and
$\psi=j_3+\alpha q$. The nondegenerate two-dimensional orbits are one-dimensional
fibrations with the fiber local coordinates $\sigma=\pi$ over the curve
$\psi-\alpha\varphi=j_3$ --- const on the torus. Because $\alpha$ is an irrational
number, this curve forms a dense winding on the torus.

In this case, the effect of the non-Hausdorff character of the orbit space is that the
quantity $j_3$, which is defined up to $2\pi(n-\alpha m)$ with $n,m \in \mathbb{Z}$,
cannot be one of three coordinates in the orbit space. The Casimir function $K_3$
represented in the canonical coordinates is infinite-valued and has the form
$K_3=j_3+2\pi(n-\alpha m)$. If we fix the value of $j_3$ at the initial moment, then
during the evolution process (determined by system (\ref{eq12})), the numbers $n$ andm
change by jumps, i.e., the values of the Casimir function $K_3$ jump from one branch to
the other. In Fig. 1, the value of the Casimir function $K_3=\arctan(P_4/\alpha
P_5)-\alpha\arctan(P_2/P_3)$ obtained numerically for the Hamiltonian
$H=(P^2_1+2P_1P_3+2P_1P_4 -P^2_4+P^2_5 )/2$ of system (\ref{eq13}) is shown as an
illustration.

We note that the ambiguity of the quantity $j_3$ is no obstacle to integrating the
geodesic flow with the central metric using canonical transformation (\ref{eq30}),
(\ref{eq31}) because this quantity only enters the functions $\sin(j_3 -\alpha q)$ and
$\cos(j_3 - \alpha q)$, and the variable $q$ is defined modulo $2\pi$.

In this case, the functions $T^m(x,p)$ are given by
$$
T^1(x,p)=\frac{x_2p_2+x_3p_3}{\sqrt{p_2^2+p_3^2}},\quad
T^2(x,p)=\frac{x_1p_1+x_4p_4}{\sqrt{p_1^2+p_4^2/\alpha^2}},\quad T^3(x,p)=\alpha x_4^2
p_1 -\frac{x_1p_4}{\alpha.}
$$
The canonical transformation $(x,p) \rightarrow (q,\pi,j,\tau)$ is therefore defined by
the expressions
$$
x_3p_2 - x_2p_3 + x_1p_4  \alpha^2 x_4p_1 = \pi,\quad p_2=j_1 \sin q,\quad p_3=j_1 \cos
q,
$$
$$
p_4=\alpha j_2 \sin(j_3+\alpha q),\quad p_1=j_2 \cos(\alpha q + j_3),
$$
$$
\frac{x_2p_2+x_3p_3}{\sqrt{p_2^2+p_3^2}}=\tau^1,\quad
\frac{x_1p_1+x_4p_4}{\sqrt{p_1^2+p_4^2/\alpha^2}}=\tau^2,\quad \alpha x_4^2
p_1-\frac{x_1p_4}{\alpha}=\tau^3.
$$
For an arbitrary matrix $\mathbf{G}^{AB}$, Hamiltonian system (\ref{eq35}) is
two-dimensional and is integrated in quadratures using the "energy" integral
$\widetilde{H}(q,\pi,j)=\mathbf{G}^{11}\pi^2/2 + A(q,j)\pi + B(q,j)/2$, where
$$
A(q,j)=j_1(\mathbf{G}^{13} \cos q + \mathbf{G}^{12} \sin q) + j_2 ( \mathbf{G}^{15}
\cos(j_3 + \alpha q) + \mathbf{G}^{14} \sin(j_3 + \alpha q));
$$
$$
B(q,j)=\frac12\, j^2_2 \alpha^2 \mathbf{G}^{44} (1 - \cos 2(j_3 + \alpha q)) +
$$
$$
+ \alpha j_1 j_2 \mathbf{G}^{34} ( \sin(j3 + \alpha q + q) + \sin(j_3 + \alpha q -  q) )+
j^2_1 \mathbf{G}^{23} \sin 2q +
$$
$$
+ \alpha j_1 j_2 \mathbf{G}^{24} (  \cos(j_3+ \alpha q - q) + \cos(j_3 + \alpha q + q) )
+ \frac12\, j^2_1 (\mathbf{G}^{22} + \mathbf{G}^{33} ) -
$$
$$
- \frac12\, j^2_1 (\mathbf{G}^{22} - \mathbf{G}^{33}) \cos 2q + \frac12\, j^2_2
\mathbf{G}^{55} ( 1 + \cos 2(j_3 + \alpha q)) +
$$
$$
+ \alpha j^2_2 \mathbf{G}^{45} \sin 2(j_3 + \alpha q) + j_1 j_2 \mathbf{G}^{25} (\sin(j_3
+ \alpha q + q) - sin(j_3 + \alpha q - q)) +
$$
$$
+ j_1 j_2 \mathbf{G}^{35} ( \cos(j_3 + \alpha q + q) + cos(j_3 +\alpha q - q)) .
$$

\end{document}